\documentclass[preprint,aps]{revtex4}

\usepackage{graphics}
\usepackage{epsfig}
\usepackage{dcolumn}
\usepackage{bm}

\begin{document}

\title{New Members in the $0^+$ ($0^{++})$ Family}

\author{Xiao-Gang He$^{1,2}$, Xue-Qian Li$^1$, Xiang Liu$^1$, Xiao-Qiang Zeng$^1$}
\affiliation{$^1$Department of Physics, Nankai
University, Tianjin}
\affiliation{$^2$NCTS/TPE, Department of
Physics, National Taiwan University, Taipei}

\date{\today}

\begin{abstract}
Recent BES data on $J/\psi \to \phi \pi\pi$ indicate that there is
a possible new spin-0 state ($f_0(1790)$) with a mass of
$m=1790^{+40}_{-30} \mathrm{MeV}/\mathrm{c}^2$. Assuming it to be
an iso-singlet $0^+(0^{++})$, we propose a new mixing scheme to
describe this and the other three known iso-singlet $f_0(1370)$,
$f_0(1500)$, $f_0(1710)$ states by adding iso-singlet hybrid
states to the usual basis of two iso-singlet quarkonia and one
glueball. Since there are two iso-singlet hybrid states, $(u\bar
u+d\bar d)g/\sqrt 2$ and $s\bar sg$, this new basis implies
existence of another iso-singlet state $X$. Using known data, we
estimate the ranges of the mixing parameters. We find two sets of
solutions with X mass predicted to be about 1820 MeV and 1760 MeV,
respectively. We also study implications on the decay properties
of these new states.
\end{abstract}


\maketitle


The BES collaboration has recently obtained evidence for a new
broad state in the spectrum of $\pi\pi$ in $J/\psi\rightarrow \phi
\pi\pi$ decay. Their results indicate that it is a $0^{+}$ state
with mass and width given by $m=1790^{+40}_{-30}
\mathrm{MeV}/\mathrm{c}^2$ and
$\Gamma=270^{+60}_{-30}\mathrm{MeV}/\mathrm{c}^2$. The observed
branching ratio for $\mathcal{B}(J/\psi\rightarrow \phi
f_{0}(1790))\cdot \mathcal{B}(f_{0}(1790)\rightarrow \pi\pi)$ is
determined to be $(6.2\pm1.4)\times 10^{-4}$ \cite{bes-1790}. This
resonant state is named as $f_{0}(1790)$.

In the energy range of 1 to 2 GeV, three $0^+(0^{++})$ states:
$f_{0}(1370)$, $f_{0}(1500)$ and $f_{0}(1710)$ have been
experimentally confirmed \cite{pdg}. The possible new state
$f_0(1790)$ may be a new member of the $0^+(0^{++})$ family, even
though its isospin and G-parity have not well determined yet.
Close et al. suggested that the three resonant states
($f_{0}(1370)$, $f_{0}(1500)$ and $f_{0}(1710)$) may be mixtures
of a scalar glueball $G$, an iso-singlet quarkonium $N=(u\bar
u+d\bar{d})/\sqrt{2}$ and  an $S=s\bar s$ \cite{close}. Several
other authors have also discussed the mixing mechanism of the
three resonant states and related phenomenology
\cite{3states,shen,close-mixing}. If the possible new $f_0(1790)$
state is another member of the $0^+(0^{++})$ family, it may also
mix with the three known states. To describe this possible new
state, it is necessary to enlarge the previously used basis in
terms of constituent quarks and gluons. A natural way of achieving
this is to introduce the $q\bar q g$ hybrid states, even though
other possibilities exist\cite{close-zhao}. We therefore propose
that the basis in terms of constituent quarks and gluons for a
unified description of states in the $0^+(0^{++})$ family is
composed of the glueball state $G$, quarkonia $N=(u\bar
u+d\bar{d})/\sqrt{2}$, $S=s\bar s$ and the two new hybrid states:
$(u\bar u+d\bar d)g/\sqrt{2}$ and $(s\bar s)g$. With this picture,
we predict existence of a yet to be discovered new physical state
$X$. Since the hybrid states have very different structure
compared with the usual glueball and quarkonia, the new states
will have some distinctive new signatures in some decays,
especially for the doubly OZI suppressed decay processes. In this
paper we study implications of this mixing mechanism.

Below 1 GeV there are also other $0^{++}$ states, such as
$f_0(600)$ and $f_0(980)$\cite{pdg}. These states have masses much
lower than that of other members of the $0^+(0^{++})$ family
mentioned above, therefore $f_0(600)$ and $f_0(980)$ can hardly
mix with the other heavier states. We will not discuss them in
this paper.

If the parameters are known, one can diagonalize the mass matrix
and obtain the eigenmasses and mixing parameters of the physical
states. The mixing parameters are, however, completely governed by
non-perturbative effects which cannot be reliably evaluated at
present. Therefore, we will use the experimental data, as much as
possible, as inputs to obtain the mixing parameters.


We now study possible structures for the mixing. The effective
Hamiltonian $\mathcal{H}$ for the system cannot be calculated from
QCD yet because of complicated non-perturbative effects. With
certain simplifications, the form of the mass matrix for the $G$,
$N$ and $S$ states has been suggested by Close et al. and some
other authors \cite{shen,close-mixing}, where $G$ can strongly
couple to both quarkonia $N$ and $S$, but the element $\langle
N|\mathcal{H}|S\rangle$ is obviously OZI suppressed and can
therefore be neglected at the lowest order approximation. Since
this coupling is flavor-independent, one has the relation
$e=\langle G|\mathcal{H}|\xi_{S}\rangle=\langle
G|\mathcal{H}|\xi_{N}\rangle/\sqrt{2}$. In analog, we assume that
only the coupling of glueball to the hybrids is strong, thus
$f=\langle G|\mathcal{H}|S\rangle=\langle
G|\mathcal{H}|N\rangle/\sqrt{2}$ is substantial while other matrix
elements can be practically set to be null. With the approximation
described here, the mass matrix can be expressed as
\begin{eqnarray}
M=\left (
\begin{array}{ccccc}
M_{\xi_{S}}&0&e&0&0\\
0&M_{\xi_{N}} & \sqrt{2}e & 0&0\\
e&\sqrt{2}e& M_{G}& f & \sqrt{2}f\\
0&0& f& M_{S}&0\\
0&0&\sqrt{2}f&0&M_{N}
\end{array} \right ),\label{matrix}
\end{eqnarray}
where $M_{\xi_{S}}=\langle \xi_S|\mathcal{H}|\xi_S\rangle$,
$M_{\xi_{N}}=\langle \xi_N|\mathcal{H}|\xi_N\rangle$,
$M_{G}=\langle G|\mathcal{H}|G\rangle$, $M_{S}=\langle
S|\mathcal{H}|S\rangle$ and $M_{N}=\langle N|\mathcal{H}|N\rangle$
are the diagonal matrix elements of $M$.

Diagonalizing the above matrix, one obtains the mass eigenvalues
and physical states in terms of the quarkonia, hybrids and
glueball. We parameterize the relation between the physical states
and the basis  as
\begin{eqnarray}
&&F_{phys} = U B_{basis},\;\;U=\left (
\begin{array}{ccccc}
v_{1}&w_{1}& z_{1} & y_{1}&x_{1}\\
v_{2}&w_{2}& z_{2} & y_{2}&x_{2}\\
v_{3}&w_{3}& z_{3} & y_{3}&x_{3}\\
v_{4}&w_{4}& z_{4} & y_{4}&x_{4}\\
v_{5}&w_{5}& z_{5} & y_{5}&x_{5}
\end{array} \right ),
\end{eqnarray}
where $F_{phys}^T = (|X\rangle$, $|f_{0}(1790)\rangle$,
$|f_{0}(1710)\rangle$, $|f_{0}(1500)\rangle$, $|f_{0}(1370))$ and
$B_{basis}^T = (|\xi_{S}\rangle,\;\; |\xi_{N}\rangle,\;\;
|G\rangle,\;\; |S\rangle,\;\; |N\rangle)$. Here the state $X$ is
an extra $0^{++}$ state predicted in this scheme.

As $\mathcal{H}$ is not derivable and therefore neither all the
matrix elements, we need to determine them by fitting data except
the scalar glueball mass $M_G$. In our later discussions we will
take the lattice calculation results of \cite{lattice} to
constrain $M_G$ to be within the range $1.5 \sim 1.7$ GeV . The
mixing parameters $v_i$, $z_i$ and $y_i$ depend on the seven
parameters $M_{\xi_S,\xi_N,G,S,N}$, $e$ and $f$. The available
data which are directly related to these parameters are the four
known eigenmasses of $f_0(1790,1710,1500,1370)$. To completely fix
all the parameters, more information is needed. To this end, we
use information from the ratios of the measured branching ratios
of $f_0(1790,1710,1500,1370)$ to two pseudoscalar mesons listed in
Table 1.

The effective Hamiltonian of scalar state decaying into two
pseudoscalar mesons can be written as \cite{lagrangian}
\begin{eqnarray} \label{decay}
\mathcal{H}^{PP}_{eff}&=&f_{1}\mathrm{Tr}[X_{F}P_{F}P_{F}]
+f_{2}X_{G}\mathrm{Tr}[P_{F}P_{F}] \nonumber\\
&+&f_{3}X_{G}\mathrm{Tr}[P_{F}]\mathrm{Tr}[P_{F}]
\nonumber+f_{4}\mathrm{Tr}[X_{H}P_{F}P_{F}]\nonumber\\
&+&
f_{5}\mathrm{Tr}[X_{H}P_{F}]\mathrm{Tr}[P_{F}]+f_{6}\mathrm{Tr}[X_{F}]\mathrm{Tr}[P_{F}P_{F}]
\nonumber\\
&+& f_{7}\mathrm{Tr}[X_{F}P_{F}]\mathrm{Tr}[P_{F}]
+f_{8}\mathrm{Tr}[X_{F}]\mathrm{Tr}[P_{F}]\mathrm{Tr}[P_{F}]\nonumber\\
&+&f_{9}\mathrm{Tr}[X_{H}]\mathrm{Tr}[P_{F}P_{F}]
+f_{10}\mathrm{Tr}[X_{H}]\mathrm{Tr}[P_{F}]\mathrm{Tr}[P_{F}].
\end{eqnarray}
Here $X_{F}$ is the flavor matrices of iso-singlet quarkonia
components of $X_{i}$ where the subscript $i=1,...,5$  labels the
five physical states. The detailed expression for $X_{F}$ is given
as \cite{shen}
\begin{eqnarray}
X_{F}&=&a\lambda^{0}+b\lambda^{8}=\left(\begin{array}{ccc} \frac{u\bar u +d\bar d}{2}&0&0\\
0&\frac{u\bar u +d\bar d}{2}&0\\
0&0&s\bar s
\end{array}\right)\nonumber\\
&=&\left (\begin{array}{ccc} \sum_{i}\frac{x_{i}}{\sqrt{2}}X_{i}&0&0\\
0&\sum_{i}\frac{x_{i}}{\sqrt{2}}X_{i}&0\\
0&0&\sum_{i}y_{i}X_{i}
\end{array}\right)
\end{eqnarray}
and $P_F$ is the pesudoscalar octet,
\begin{eqnarray}
P_{F}=\left(\begin{array}{ccc}
\frac{\pi^{0}}{\sqrt{2}}+\frac{x_{\eta}\eta+x_{\eta'}\eta'}{\sqrt{2}}&\pi^{+}&K^{+}\\
\pi^{-}&-\frac{\pi^{0}}{\sqrt{2}}+\frac{x_{\eta}\eta+x_{\eta'}\eta'}{\sqrt{2}}&
K^{0}\\
K^- &\bar{K}^{0}&y_{\eta}\eta+y_{\eta'}\eta'
\end{array}\right).
\end{eqnarray}
In the above, $x_{\eta,\eta'}$ and $y_{\eta,\eta'}$ describe the
$\eta-\eta's$ mixing, and
\begin{eqnarray}
&&x_{\eta}=y_{\eta'}=\frac{\cos\theta-\sqrt{2}\sin\theta}{\sqrt{3}},\nonumber\\
&&x_{\eta'}=-y_{\eta}=\frac{\sin\theta+\sqrt{2}\cos\theta}{\sqrt{3}},
\end{eqnarray}
where $\theta=-19.1^\circ$\cite{etamixing} is the mixing angle of
$\eta$ and $\eta'$.

The concrete expressions of $X_{G}$ and $X_{H}$ are
\begin{eqnarray}
X_{G}&=&\sum_{i}z_{i}X_{i},\\
X_{H}&=&(a\lambda^0 +b\lambda^{8})g=\left(\begin{array}{ccc} \frac{u\bar u +d\bar d}{2}g&0&0\\
0&\frac{u\bar u +d\bar d}{2}g&0\\
0&0&s\bar sg
\end{array}\right) \nonumber\\
&=&\left(\begin{array}{ccc} \sum_{i}\frac{w_{i}}{\sqrt{2}}X_{i}&0&0\\
0&\sum_{i}\frac{w_{i}}{\sqrt{2}}X_{i}&0\\
0&0&\sum_{i}v_{i}X_{i}
\end{array}\right).
\end{eqnarray}

The $f_{6-10}$ terms in the above effective Hamiltonian describing
the decay modes with two-meson final states are OZI suppressed as
can be seen from Figure 1((6)-(10)). The contributions from these
terms can be neglected to a good approximation. Within this
approximation, 5 parameters (actually 4 parameters
$\xi_{i}=f_{1+i}/f_1$ when considering ratios of branching ratios)
are needed to describe decay modes with two pseudoscalar mesons in
the final states. We obtain the decay width $\Gamma(X_i\to \pi\pi,
K\bar K, \eta\eta, \eta\eta')$ in terms of the parameters
$\xi_{1}=f_{2}/f_{1}$, $\xi_{2}=f_{3}/f_{1}$,
$\xi_{3}=f_{4}/f_{1}$ and $\xi_{4}=f_{5}/f_{1}$.

Using the above mass matrix and the decay amplitudes, our task is
now reduced to see if the four eigenmasses of
$f_0(1370,1500,1710,1790)$, and eight ratios of the branching
ratios listed in Table 1 can be described by the 11 parameters (7
parameters in the mass matrix with $M_G$ in the range of 1.4 to
1.7 GeV plus the 4 parameters $\xi_i$ in the decay amplitudes) in
some reasonable ranges. This by no means is a trivial task. We,
however, do find parameter spaces which can give reasonable fit to
experimental data.
 Since the mass of $X$ is not
known, we consider two types of solutions for the mass $M_{X}$ of
$X$: (a) $M_{X} > M_{f_{0}(1790)}$, and (b) $M_{X} <
M_{f_{0}(1790)}$.  We display the results in the following.

 \noindent Case (a) $M_{X}>M_{f_{0}(1790)}$

In Table 2, we list the input parameters and the resulting
eigenmasses and branching ratios obtained. Our procedure to obtain
the fitted values for the input parameters is guided by obtaining
numbers which are mostly consistent with the central values of the
known data as the parameters being within the reasonable ranges
described earlier. The mixing matrix is given by
\begin{eqnarray*}
U = \left ( \begin{array}{lllll}
-0.986&-0.107&-0.109&-0.065&-0.030\\
-0.131&+0.972&+0.146&+0.121&+0.045\\
-0.077&-0.173&+0.273&+0.937&+0.105\\
-0.061&-0.099&+0.617&-0.284&+0.725\\
+0.041&+0.063&-0.715&+0.147&+0.679 \end{array}
\right ).
\end{eqnarray*}
The dominant component of $f_{0}(1790)$ is
$(u\bar{u}+d\bar{d})g/\sqrt{2}$, whereas  $s\bar{s}g$ is the
dominant one in $X$. The main components of $f_{0}(1710)$, and
$f_{0}(1500,1370)$ are S and mixtures of N and G, respectively.
The mass of $X$ is approximately 1.823 GeV.

\noindent{Case (b) $M_{X} < M_{f_{0}(1790)}$}

There are some differences for case (b) from case (a). The input
parameters, the resulting eigenmasses and branching ratios
obtained are also listed in Table 2. The mixing pattern is given
by
\begin{eqnarray*}
U = \left ( \begin{array}{lllll}
-0.168&+0.945&+0.179&+0.210&+0.060\\
-0.978&-0.125&-0.129&-0.099&-0.039\\
-0.095&-0.274&+0.248&+0.919&+0.096\\
-0.067&-0.111&+0.614&-0.282&+0.725\\
+0.044&+0.068&-0.716&+0.147&+0.678\end{array}\right ).
\end{eqnarray*}
In this case, the dominant component of $f_{0}(1790)$ is
$s\bar{s}g$, whereas the $(u\bar{u}+d\bar{d})g/\sqrt{2}$ is the
dominant one in $X$. The main components of $f_{0}(1710)$, and
$f_{0}(1500, 1370)$ still are, respectively, S and mixitures of N
and G. The mass of $X$ is approximately 1.76 GeV in this case. We
note that the ratio $B(f_0(1790)\to \pi\pi)/B(f_0(1790)\to K\bar
K)$ in this case is below the central value of the data
($B(f_0(1790)\to \pi\pi)/B(f_0(1790)\to K\bar K)=3.88$). However,
due to large error associated with the data, this case cannot be
ruled out at present. This may provide a crucial criteria to
distinguish the two cases.


There are many possible decay modes for the  new state
$f_{0}(1790)$ and $X$. We will consider several two-body decay
modes of these states. They are $f_{0}(1790)(X)$ decays into two
pseudoscalar-mesons, two vector-mesons and two-photons.

\noindent {\bf $f_{0}(1790) (X) \to P P'$}

The two-pseudoscalar meson decays have been given in
eq.(\ref{decay}) previously. We have used several of these decay
modes involving $f_0(1370,1500,1710,1790)$ to fix the parameters.
Using the mixing parameters determined, the decay modes for
$f_{0}(1790)$ and $X$ can be predicted. We obtain the results for
cases (a) and (b) in the following.

For case (a), we have
\begin{eqnarray*}
&&B(f_{0}(1790) \to \pi\pi): B(f_{0}(1790)\to K \bar
K)\nonumber\\
&&: B(f_{0}(1790) \to
\eta\eta): B(f_{0}(1790)\to \eta \eta')\nonumber\\
&&= 23:10:5:2,\nonumber \\
&&B(X\to \pi\pi):B(X\to K \bar K): B(X \to \eta\eta): B(X\to
\eta \eta')\nonumber \\
&&= 5:43:6:4,
\end{eqnarray*}
whereas for case (b), we have
\begin{eqnarray*}
&&B(f_{0}(1790) \to \pi\pi): B(f_{0}(1790)\to K \bar
K)\nonumber\\&&: B(f_{0}(1790) \to
\eta\eta): B(f_{0}(1790)\to \eta \eta')\nonumber\\&& = 13:31:6:0.3, \\
&&B(X\to \pi\pi):B(X\to K \bar K): B(X \to \eta\eta): B(X\to \eta
\eta')\\&& =10:15:5:0.4.
\end{eqnarray*}

Using the above obtained ratios of $\Gamma(f_{0}(1790)(X)\to
PP')/\Gamma(f_0(1710)\to K\bar K)$ and combining the measured
value of $\Gamma(f_0(1710)\to K\bar K)$, we obtain the
corresponding values for $f_{0}(1790)(X)\to PP'$ in our Table 3.
Since the total width of $f_{0}(1790)$ is measured at BES, one can
obtain branching ratios of $f_{0}(1790)\to PP'$, by contraries,
for $X$, one can only have the partial widths.

\noindent {\bf $f_{0}(1790) (X) \to VV'$}

Now let us turn to the case of decays with two vector mesons $VV'$
in the final state. The effective Hamiltonian is similar to that
for the pseudoscalar-meson case in eq.(\ref{decay}). One just
replace the $P_FP_F$ by $\partial^\mu V^\nu \partial_\nu V_\mu$ at
appropriate places with $V$ being the vector nonet.

Since the measurements on such $VV'$ channels, even for the
confirmed resonant states $f_{0}(1370)$, $f_{0}(1500)$ and
$f_{0}(1710)$ are absent, it is not possible to make a definite
evaluation on their branching ratios yet, and even the ratios
among the branching ratios. But these decay modes should occur
with substantial branching ratios. We will come back to this
later.

\noindent {\bf $f_{0}(1790)(X) \to \gamma \gamma$}

We now consider the two-photon decay modes.  In the spirit of
Ref.\cite{gamma} that the decay amplitude is proportional to the
electric charge coupling of the two photon at the quark level,
ignoring mass-dependent effects, we obtain the following.

For case (a)
\begin{eqnarray}
&&\Gamma(X\rightarrow \gamma\gamma):\Gamma(f_{0}(1790)\rightarrow
\gamma\gamma):\Gamma(f_{0}(1710)\rightarrow
\gamma\gamma)\nonumber\\&&:\Gamma(f_{0}(1500)\rightarrow
\gamma\gamma):\Gamma(f_{0}(1370)\rightarrow \gamma\gamma)
\nonumber\\
&& =0.06:0.16:3.42:10.39:12.98.
\label{gamma-gamma}
\end{eqnarray}

For case (b)
\begin{eqnarray}
&&\Gamma(X\rightarrow \gamma\gamma):\Gamma(f_{0}(1790)\rightarrow
\gamma\gamma):\Gamma(f_{0}(1710)\rightarrow
\gamma\gamma)\nonumber\\&&:\Gamma(f_{0}(1500)\rightarrow
\gamma\gamma):\Gamma(f_{0}(1370)\rightarrow \gamma\gamma)
\nonumber\\&& =0.36:0.11:3.17:10.41:12.94.\label{gamma-gamma1}
\end{eqnarray}


We have proposed a mixing scheme of isosinglet $0^+(0^{++})$
states: quarkonia, glueball and hybrid state $q\bar q g$ to
accommodate a possible new state $f_0(1790)$ and other known
states in the $0^+$ $(0^{++})$ family with masses in the range
between 1 to 2 GeV. Using known experimental data, we have been
able to obtain information on the mixing parameters. The related
phenomenology indicates that the parameters obtained from this
fitting are within reasonable ranges. If the $f_0(1790)$ state is
confirmed, this scheme predicts the existence of a new particle
$X$. This can be tested further with BES data.

The decay channels of $f_0(1790)$ ad $X$ to two pseudoscalar
mesons can provide important information about the mixing
mechanism and distinguish cases (a) and (b). For case (a) all
possible decay modes have large branching ratios which may be
measured by improved experiments, whereas for case (b), the decays
of $f_0(1790)$ to $\eta\eta$, $\eta\eta'$, and $X$ to $\eta\eta'$
have substantially smaller branching ratios. Particularly the
ratio for $r=B(f_0(1790)\to \pi\pi)/B(f_0(1790)\to K\bar K)$ is a
very important criterion to test which case is more realistic
since for case (a) this ratio is about 2 and for case (b) it is
about 0.4. This can be easily understood by noticing that the main
component of $f_0(1790)$ in case (a) is $(u\bar u+d\bar d)g/\sqrt
2$ which has a much larger probability to transit into $\pi\pi$
compared with case (b) where the main component of $f_0(1790)$ is
$s\bar sg$. Therefore the central value ($r=3.88$) of the present
data favors case (a) over case (b) although a definitive
conclusion cannot be drawn at this stage due to larger
experimental errors.

Another interesting fact to note is that\cite{jins} $f_0(1710)$ is
observed in the $K \bar K$ spectrum in $J/\psi\rightarrow \omega
K\bar K$ and $J/\psi\rightarrow \phi K\bar K$ decays, but not in
$J/\psi\rightarrow \phi \pi\pi $. There has been some theoretical
effort to explain this observation \cite{glozman}. In our picture
this is also very natural since the main component of $f_0(1710)$
is $s\bar s$ which does not directly transit to $\pi\pi$, so that
$f_0(1710)\rightarrow \pi\pi$ would be much suppressed compared
with $f_0(1710)\to K\bar K$.

Obviously, by contraries, the radiative decay modes to
$\gamma\gamma$ would be very difficult to measure via $J/\psi\to
\gamma f_0(1790)(X)\to \gamma\gamma\gamma$ , because the final
states with only three photons are hard to be reconstructed. We
hope that our experimental colleagues can figure out some ways to
make the difficult measurements.

With the present data, we are not able to make detailed
predictions for the decay modes with two vector mesons in the
final state. However, from the diagrams shown in Fig.1, we can
expect that some decay modes may be measured and help us to gain
more information about the properties of the new states
$f_0(1790)$ and $X$. The radiative decays such as $J/\psi \to
\gamma f_0(1790)\to \gamma V V'$ and $J/\psi \to \gamma X\to
\gamma VV'$ are promising channels to study $f_0(1790)$ and $X$
states. A particularly interesting channel is $J/\psi \to \gamma
\phi \omega$ since this is a doubly OZI suppressed processes if
there is not a hybrid intermediate state. With a hybrid state $X$,
the process $J/\psi \to \gamma X\to \gamma \phi \omega$ may have a
large branching ratio. This may happen for case (a) since $X$ has
a mass about 1820 MeV, but not possible for case (b) since in this
case the $X$ mass of 1760 MeV is below the threshold. We strongly
urge our experimental colleagues to carry out precise measurements
to
test the mechanism proposed here.\\

\noindent {\bf Acknowledgements}: We thank Dr. S. Jin and Dr. X.
Shen for discussions.  This work is partly supported by
NNSFC and NSC.\\

\noindent {\bf Note Added} When we were ready to submit this
paper, we saw a paper on the archive by B. A. Li
(hep-ph/0602072)\cite{li} who cited that the BES collaboration
reported\cite{bbbes} the observation of a state X(1810) in the
spectrum of $\omega \phi$ in $J/\psi\rightarrow \gamma \omega
\phi$. This newly observed  state fits our prediction of case (a)
well. B. A. Li proposed the $X(1810)$ to be a four-quark state
which is different from our hybrid state description. After
submitting the paper on the archive we also became aware of the
papers\cite{vvv} by Vijande et al. who discussed $f_0(1790)$ in
the framework of mixing of a chiral nonet tetraquarks with
conventional $q\bar q$ states. We thank A. Valcarce for bring this
paper to our attention.

\newpage

\begin{table}[htb]
\begin{center}
\begin{tabular}{|c|c|c|c|c|} \hline
&Experiment \cite{WA}& (a) Fitted & (b) Fitted\\
\hline $\frac{\Gamma(f_{0}(1370)\rightarrow
\pi\pi)}{\Gamma(f_{0}(1370)\rightarrow K\bar{K})}$& $2.17\pm0.90$&0.55&0.47\\
\hline $\frac{\Gamma(f_{0}(1370)\rightarrow
\eta\eta)}{\Gamma(f_{0}(1370)\rightarrow K\bar{K})}$& $0.35\pm0.30$ &0.32&0.34\\
\hline $\frac{\Gamma(f_{0}(1500)\rightarrow
\pi\pi)}{\Gamma(f_{0}(1500)\rightarrow \eta\eta)}$&$5.56\pm0.93$  &5.07 &5.18\\
\hline $\frac{\Gamma(f_{0}(1500)\rightarrow
K\bar{K})}{\Gamma(f_{0}(1500)\rightarrow \pi\pi)}$& $0.33\pm0.07$& 0.48 &0.45\\
\hline $\frac{\Gamma(f_{0}(1500)\rightarrow
\eta\eta')}{\Gamma(f_{0}(1500)\rightarrow \eta\eta)}$& $0.53\pm0.23$ &0.09 &0.10\\
\hline $\frac{\Gamma(f_{0}(1710)\rightarrow
\pi\pi)}{\Gamma(f_{0}(1710)\rightarrow K\bar{K})}$&$0.20\pm0.03$  & 0.16 &0.17\\
\hline $\frac{\Gamma(f_{0}(1710)\rightarrow
\eta\eta)}{\Gamma(f_{0}(1710)\rightarrow K\bar{K})}$&
$0.48\pm0.19$ & 0.19 &0.19
\\\hline
$\frac{\Gamma(f_{0}(1790)\rightarrow
\pi\pi)}{\Gamma(f_{0}(1790)\rightarrow K\bar{K})}$&$3.88^{+5.6}_{-1.9}$\cite{bes-1790} &2.22 &0.42\\
\hline
\end{tabular}
\end{center}
\caption{The measured and predicted central values for branching
ratios.}\label{fig}
\end{table}

\begin{table}[htb]
\begin{center}
\begin{tabular}{|c|c|c|c|c|c|} \hline
Parameter & (a) Fitted & (b) Fitted & Parameter & (a) Fitted& (b)
Fitted\\\hline
$M_{H_{S}}$(GeV)&1.82& 1.79&e(GeV)&0.03&0.03\\
$M_{H_{N}}$(GeV)&1.78&1.75 &f(GeV)&0.08&0.08\\
$M_{G}$(GeV)&1.43& 1.43&$\xi_{1}$&1.27&1.46\\
$M_{S}$(GeV)&1.69& 1.69&$\xi_{2}$&0.41&0.40\\
$M_{N}$(GeV)&1.42& 1.42&$\xi_{3}$&0.80&0.10\\
$M_{X}$(GeV)&1.823&1.758 &$\xi_{4}$& 0.10&0.10\\
$M_{f_{0}(1790)}$(GeV)&1.786&1.794&&&\\
$M_{f_{0}(1710)}$(GeV)&1.713&1.712&&&\\
$M_{f_{0}(1500)}$(GeV)&1.516&1.516&&&\\
$M_{f_{0}(1370)}$(GeV)&1.301&1.301&&&\\\hline
\end{tabular}
\end{center}
\caption{The values for the parameters in the mass matrix and the
$PP$ decay amplitudes.}
\end{table}

\begin{table}[htb]
\begin{center}
\begin{tabular}{|c|c|c|} \hline
&(a)&(b)\\\hline
$BR(f_{0}(1790)\rightarrow\pi\pi)$&$23.0\%$&$1.3\%$\\
$BR(f_{0}(1790)\rightarrow K\bar{K})$&$10.3\%$&$3.1\%$\\
$BR(f_{0}(1790)\rightarrow\eta\eta)$&$4.5\%$&$0.6\%$\\
$BR(f_{0}(1790)\rightarrow\eta\eta')$&$2.3\%$&$0.03\%$\\
$\Gamma(X\rightarrow\pi\pi)$ MeV&$5.2$&10.2\\
$\Gamma(X\rightarrow K\bar{K})$ MeV&$44.8$&14.8\\
$\Gamma(X\rightarrow\eta\eta)$ MeV &$6.8$&5.4\\
$\Gamma(X\rightarrow\eta\eta')$ MeV&$4.6$&0.4\\\hline
\end{tabular}\label{fig-2}
\end{center}
\caption{The branching ratios of $f_{0}(1790)\to PP'$ and the
widths of  $X\to PP'$.}
\end{table}

\begin{figure}[htb]
\begin{center}
\begin{tabular}{ccccc}
\scalebox{0.4}{\includegraphics{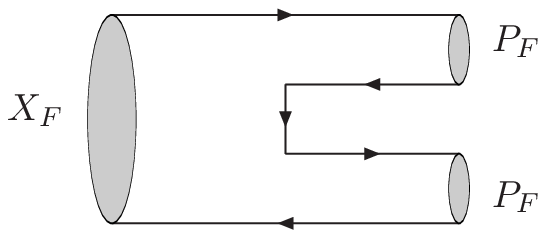}}&\scalebox{0.4}{\includegraphics{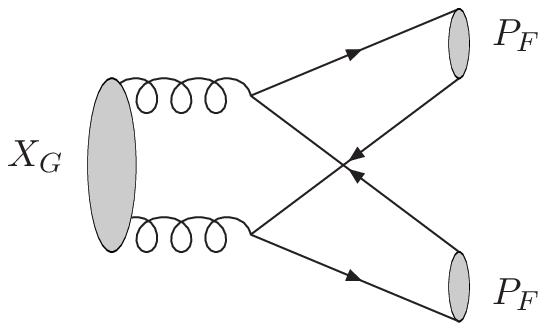}}
&\scalebox{0.4}{\includegraphics{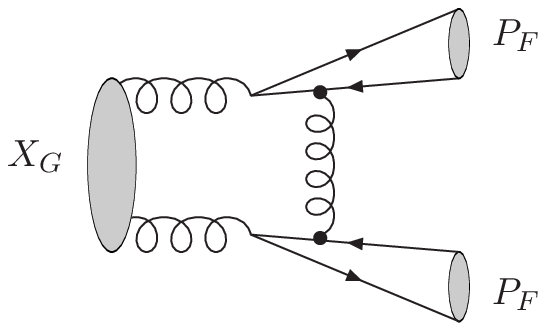}}
&\scalebox{0.4}{\includegraphics{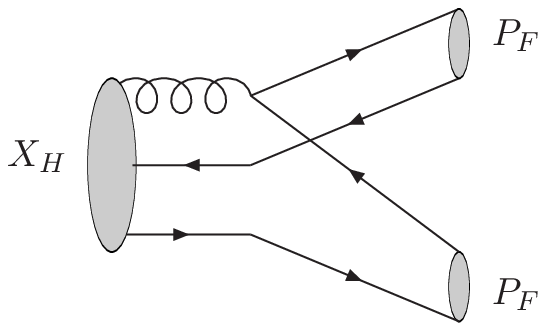}}&\scalebox{0.4}{\includegraphics{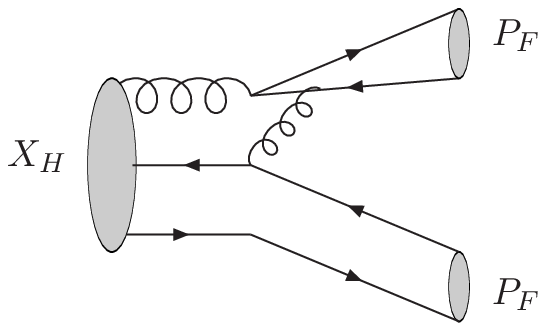}}
\\(1)&(2)&(3)&(4)&(5)\\
\scalebox{0.4}{\includegraphics{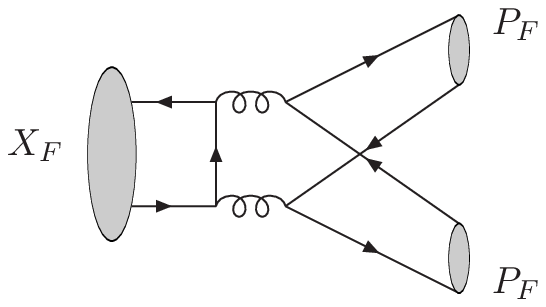}}&\scalebox{0.4}{\includegraphics{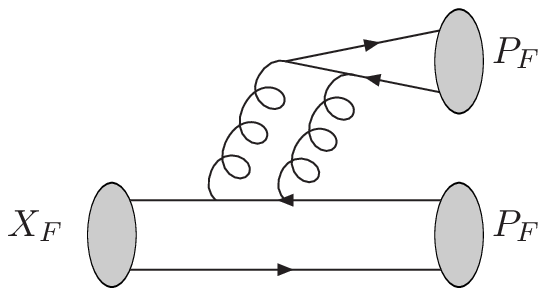}}
&\scalebox{0.4}{\includegraphics{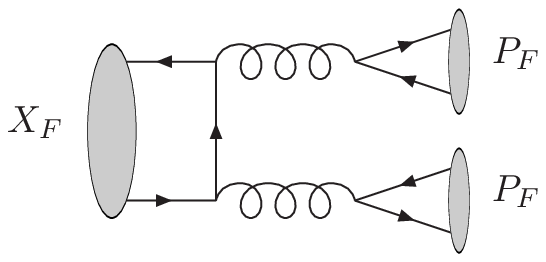}} &
\scalebox{0.4}{\includegraphics{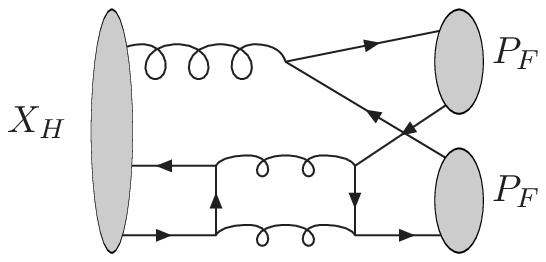}}&\scalebox{0.4}{\includegraphics{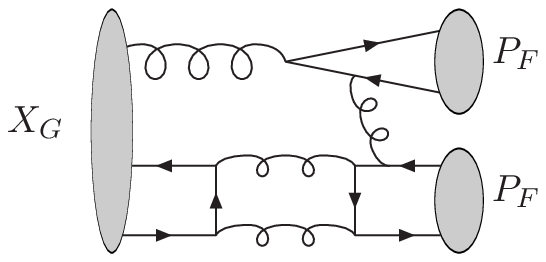}}
\\(6)&(7)&(8)&(9)&(10)\\
\end{tabular}
\end{center}
\caption{The diagrams correspond respectively to terms in
eq.(\ref{decay}). The last five terms are OZI suppressed ones.
}
\end{figure}

\end{document}